\documentclass[11pt,eqs]{article}
\usepackage{latexsym}
\usepackage{amsmath}
\usepackage{hyperref}
\hypersetup{
	colorlinks=true,
	linkcolor=cyan,
	filecolor=magenta,      
	urlcolor=cyan,
}

\makeatletter
\newcommand\xleftrightarrow[2][]{%
	\ext@arrow 9999{\longleftrightarrowfill@}{#1}{#2}}
\newcommand\longleftrightarrowfill@{%
	\arrowfill@\leftarrow\relbar\rightarrow}
\makeatother

\title{
	Noncommutativity and nonassociativity of type II superstring with coordinate dependent RR field
	\thanks{Work supported in part by the Serbian Ministry of Education,Science and Technological Development.}}
\author{B. Nikoli\'c, D. Obri\'c and B. Sazdovi\'c \thanks{email:bnikolic, dobric, sazdovic@ipb.ac.rs}\\{\it Institute of Physics Belgrade, University of Belgrade, Pregrevica 118, Serbia} }

\begin{document}
	
	\maketitle
	
\begin{abstract}

In this paper we will consider noncommutativity that arises from bosonic T-dualization of type II superstring in presence of Ramond-Ramond (RR) field, which linearly depends on the bosonic coordinates $x^\mu$. The derivative of the RR field $C^{\alpha\beta}_\mu$ is infinitesimal. We will employ generalized Buscher procedure that can be applied to cases that have coordinate dependent background fields. Bosonic part of newly obtained T-dual theory is non-local. It is defined in non-geometric space spanned by Lagrange multipliers $y_\mu$. We will apply generalized Buscher procedure once more on T-dual theory and prove that original theory can  be salvaged. Finally, we will use T-dual transformation laws along with Poisson brackets of original theory to derive Poisson bracket structure of T-dual theory and nonassociativity relation. Noncommutativity parameter depends on the supercoordinates $x^\mu$, $\theta^\alpha$ and $\bar\theta^\alpha$, while nonassociativity parameter is a constant tensor containing infinitesimal $C^{\alpha\beta}_\mu$.

\end{abstract}

\section{Introduction}
\setcounter{equation}{0}

In 1982 emerged a model \cite{original paper} that would offer the possibility of obtaining bosonic coordinates of space-time as emergent properties of more fundamental fermionic coordinates. While this model worked with supersymmetric particle, this approach suggested that maybe we can express bosonic coordinates as a Poisson bracket of fermionic coordinates. In addition to these Poisson brackets, Poisson brackets between bosonic coordinates as well as between bosonic and fermionic coordinates would remain zero. Later, in paper \cite{original paper 2}, it has been suggested that the same result can be obtained in context of string theory in case of coordinate dependent RR field. It is also suggested that unlike supersymmetric particle model, coordinate dependent RR field would produce full spectrum of non-commutative relations, bosonic coordinates themselves would become non-commutative. In this paper our goal is to determine if coordinate dependent RR field, while remaining background fields are as simple as consistency relations allow, can produce suggested non-commutative relations.

Superstring theory, as a theory of extended objects propagating in space-time, is defined in 10 dimensions \cite{book1}\cite{book2}. In order to establish link between this mathematical model and real world observations, surplus space-like dimensions are compactified on circles of radius $R$. From this kind of geometry arises new kind of symmetry, T-duality, that links theories that have radii of compactification $R$ with ones that have radii of compactification $\alpha^\prime/R$ \cite{T-duality explained 1}\cite{T-duality explained 2}. Existence of T-duality between different theories implies that those theories are physically equivalent and it gives us a way to explore how geometry and topology of one theory is connected to other. This connection between different geometries makes T-duality a useful tool in examining emergence of non-commutativity in context of closed strings \cite{Luest T-duality and geometry}.

While in string theory both open and closed strings, under certain conditions, exhibit emergence of non-commutativity, mechanisms that enable this emergence are different. In case of open string, we have that endpoints of string that propagates in presence of constant metric and Kalb-Ramond field become non-commutative \cite{open string non-commutativity}. Basic idea of open string non-commutativity is that initial coordinates can be expressed as linear combination  of effective coordinates and momenta by employing boundary conditions. In case of closed string, we do not have string endpoints therefore we don't have emergence of non-commutativity when string propagates in presence of constant background fields. In order to achieve same effect as in case of open string, we have to use coordinate dependent background fields. By finding T-dual of theories with this kind of geometry we obtain T-dual theory in non-geometric background, where  T-dual coordinates are expressed as linear combinations of original coordinates and their conjugated momenta.

Mathematical framework for obtaining T-dual theories is standard Buscher procedure \cite{T-duality procedure 1}\cite{T-duality procedure 2}. Procedure is based on existence of shift symmetry in relevant action and its implementation can be summarized in few steps. First step is localization of translational symmetry by introduction of covariant derivatives and introduction of Lagrange multipliers that make newly introduced gauge fields nonphysical. By gauge fixing  and finding equations of motion of both gauge fields and Lagrange multipliers we obtain T-dual transformation laws. These transformation laws inserted into gauge fixed action produce T-dual action. For cases where we have coordinate dependent background fields there exist generalized Buscher procedure \cite{Generalized Buscher procedure}\cite{Gereralized Buscher procedure 2}\cite{Generalized Buscher procedure 3}\cite{Additional ref 1}\cite{Additional ref 2}, this extension has one additional step, replacement of all initial coordinates with invariant coordinates. Further extension of generalized Buscher procedure is possible\cite{Auxiliary action Buscher procedure} and it is applicable to theories that do not posses shift symmetry.

In this article we will deal with closed superstring propagating in presence of linearly coordinate dependent Ramond-Ramond (RR) field using type II superstring model in pure spinor formulation. All calculations we will do in approximation of diluted flux, which means that in all calculations we keep constant and linear terms
in infinitesimal derivative of the RR field strength. Rest of the fields, metric, Kalb-Ramond and gravitino fields are constant. Furthermore all dependence of background field on fermionic terms will be neglected for mathematical simplicity. This choice of field configuration is in full accordance with consistency equations for background fields \cite{Vertex operators}.

Because we are currently only interested in non-commutative relations between bosonic coordinates, T-dualization procedure will be applied only on bosonic part of action. To find T-dual action and T-dual transformation laws we will employ extension of generalized Buscher procedure that works with coordinate dependent background fields \cite{Generalized Buscher procedure}. After finding T-dual theory, we will apply Buscher procedure once more to see if we can obtain original theory. 

%Reason for this is that background field configuration that we selected, with no dependency on fermionic coordinates, ensures that fermionic structure of T-dual theory will not produce any non-commutative effect.

Transformation laws that connect variables from initial with variables from T-dual theory will be written in canonical form, where initial momenta are expressed in terms of the T-dual coordinates. By inverting these transformation laws we obtain how sigma derivatives of T-dual theory depend on linear combinations of coordinates and momenta of original theory. Taking into account that original theory is geometrical, both locally and globally, we have that its coordinate and conjugated momenta satisfy standard Poisson brackets. By using this fact we are able to find Poisson structure of sigma derivatives of T-dual coordinates and by doing integration, Poisson structure of T-dual coordinates is obtained.

The form of obtained non-commutativity is such that non-commutativity exists when arguments are different, $\sigma \neq \bar{\sigma}$. Imposing trivial winding conditions, we obtain string winding numbers from Poisson brackets. 

In the end, we give conclusions and in appendix we present some technical details regarding derivation of $\beta^\pm_\mu$ functions.

\section{{\bf General} type II superstring action and  {\bf choice} of background fields}
\setcounter{equation}{0}

Starting point of this investigation will be action of type II superstring theory in pure spinor formulation \cite{pure spinor formalism papers 1, pure spinor formalism papers 2, pure spinor formalism papers 3, pure spinor formalism papers 4}. We will present and explain assumed approximations in order to obtain type II pure spinor action with non-constant RR field-strength. It turns out that ghost fields are neglected and only quadratic terms are considered. Final form of this kind of action will be used in subsequent sections.

\subsection{General form of the pure spinor type II superstring action}

Sigma model of type IIB superstring has the following form \cite{Vertex operators}
\begin{equation}\label{eq:dejstvo}
S=S_0 + V_{SG}.
\end{equation}
This general form of action is expressed as a sum of the part that describes the motion of string in flat background 
\begin{equation} \label{S_0 action}
S_0 = \int_{\Sigma}d^2 \xi \left( \frac{\kappa}{2} \eta_{ \mu \nu } \partial_m x^\mu \partial_n x^\nu \eta^{m n} - \pi_\alpha \partial_- \theta^\alpha + \partial_+ \bar{\theta}^\alpha \bar{\pi}_\alpha \right) 
+S_\lambda +S_{ \bar{\lambda} }\, ,
\end{equation}
and part that governs the modifications to the background fields
\begin{equation}
V_{SG} =  \int_{\Sigma}d^2 \xi (X^T)^M A_{MN} \bar{X}^N.
\end{equation}
Modifications to the flat background are introduced by integrated form of massless type II supergravity vertex operator $V_{SG}$. The terms $S_{\lambda}$ and $S_{\bar{\lambda}}$ in (\ref{S_0 action}) are {free-field} actions for pure spinors
\begin{equation}
S_\lambda = \int_{\Sigma} d^2 \xi \omega_\alpha \partial_- \lambda^\alpha, \quad S_{\bar{\lambda}} = \int_{\Sigma} d^2 \xi \bar{\omega}_\alpha \partial_+ \bar{\lambda}^\alpha\, .
\end{equation}
Here, $\lambda^\alpha$ and $\bar{\lambda}^\alpha$ are pure spinors whose canonically conjugated momenta are $\omega_\alpha$ and $\bar{\omega}_\alpha$, respectively. Pure spinors satisfy pure spinor constraints
\begin{equation}
\lambda^\alpha (\Gamma^\mu)_{\alpha \beta} \lambda^\beta =  \bar{\lambda} (\Gamma^\mu)_{\alpha \beta} \bar{\lambda}^\beta = 0\, .
\end{equation}

In general case, vectors $X^M$ and $X^N$ as well as a supermatrix $A_{MN}$ are given by

\begin{equation}
X^M = \left(\begin{matrix}
\partial_+ \theta^\alpha\\
\Pi_+^\mu\\
d_\alpha\\
\frac{1}{2} N_+^{\mu \nu}
\end{matrix} \right), 
\quad \bar{X}^M = \left( \begin{matrix}
\partial_- \bar{\theta}^\lambda\\
\Pi_-^\mu\\
\bar{d}_\lambda\\
\frac{1}{2} \bar{N}_-^{\mu \nu}
\end{matrix} \right), \quad
A_{MN} = \begin{bmatrix}
A_{\alpha \beta} & A_{\alpha \nu} & {E_\alpha}^\beta & \Omega_{\alpha, \mu \nu} \\
A_{\mu \beta} & A_{\mu \nu} & \bar{E}_\mu^\beta & \Omega_{\mu, \nu \rho}\\
{E^\alpha}_\beta & E_\nu^\alpha & P^{\alpha \beta} & {C^\alpha}{}_{\mu \nu}\\
\Omega_{\mu \nu, \beta} & \Omega_{\mu \nu, \rho} & {\bar{C}^{\beta}{}_{\mu \nu}} & S_{\mu \nu, \rho \sigma}
\end{bmatrix},
\end{equation} 
where notation is in accordance with Ref\cite{Vertex operators}. The components of matrix $A_{MN}$ are generally functions of $x^\mu$, $\theta^\alpha$ and $\bar{\theta}^\alpha$. Components themselves are derived as expansions in powers of $\theta^\alpha$ and $\bar{\theta}^\alpha$ (for details consult \cite{Vertex operators}). The superfields $A_{\mu \nu}$, $\bar{E}_\mu^\alpha$, $E_\mu^\alpha$ and $P^{\alpha \beta}$ are known as physical superfields, while superfields that are in the first row and the first column are known as auxiliary because they can be expressed in terms of physical ones \cite{Vertex operators}. Remaining superfields $\Omega_{\mu, \nu \rho}\  (\Omega_{\mu \nu, \rho})$, $C^\alpha{}_{\mu \nu}\  ({\bar{C}^\beta{}_{\mu\nu}})$ and $S_{\mu\nu,\rho\sigma}$, are curvatures (field strengths) for physical fields.   Components of vectors $X^M$ and $\bar{X}^N$ are defined as
\begin{equation}
\Pi_+^\mu = \partial_+ x^\mu + \frac{1}{2}\theta^\alpha (\Gamma^\mu)_{\alpha \beta}\partial_+\theta^\beta, \quad  
%%%%%%%%%%%%%%%%%%%%%%%%%%%%%%%%%%%%%%%%%%%%%%%%%%%%%%%%%%%%%%%%%%%%%%%%%%%%%%%%%%%%%%%%%%%%%%%%%%%%%%%%%%%%%%%%%%%%%%%%%%%%%%%%%%
\Pi_-^\mu = \partial_- x^\mu + \frac{1}{2} \bar{\theta}^\alpha (\Gamma^\mu)_{\alpha \beta} \partial_- \bar{\theta}^\beta,
\end{equation}

\begin{align}
d_\alpha &= \pi_\alpha - \frac{1}{2} (\Gamma_\mu \theta)_\alpha \left[ \partial_+ x^\mu +\frac{1}{4} (\theta \Gamma^\mu \partial_+ \theta)
\right],  \nonumber\\
%%%%%%%%%%%%%%%%%%%%%%%%%%%%%%%%%%%%%%%%%%%%%%%%%%%%%%%%%%%%%%%%%%%%%%%%%%%%%%%%%%%%%%%%%%%%%%%%%%%%%%%%%%%%%%%%%%%%%%%%%%%%%%%%%%
\bar{d}_\alpha &= \bar{\pi}_\alpha - \frac{1}{2}(\Gamma_\mu \bar{\theta})_\alpha \left[ \partial_- x^\mu +\frac{1}{4}(\bar{\theta} \Gamma^\mu \partial_- \bar{\theta} )
\right],
\end{align}

\begin{equation}
N_+^{\mu \nu} = \frac{1}{2} \omega_\alpha {( \Gamma^{[\mu\nu]}
	)^\alpha}_\beta \lambda ^\beta, \quad \bar{N}_-^{\mu\nu} = \frac{1}{2} \bar{\omega}_\alpha {(  \Gamma^{ [\mu\nu ]  }     )^\alpha}_\beta \bar{\lambda}^\beta.
\end{equation}

The world sheet $\Sigma$ is parameterized by $\xi^m = (\xi^0 = \tau, \xi^1 = \sigma)$ and world sheet light-cone partial derivatives are defines as $\partial_\pm = \partial_\tau \pm \partial_\sigma$. Superspace in which string propagates is spanned both by bosonic $x^\mu \  (\mu = 0,1,...,9)$ and fermionic $\theta^\alpha,\  \bar{\theta}^\alpha \  (\alpha = 1,2,...,16)$ coordinates. Variables $\pi_\alpha$ and $\bar{\pi}_\alpha$ represent canonically conjugated momenta of fermionic coordinates $\theta^\alpha$ and $\bar{\theta}^\alpha$, respectively. Fermionic coordinates and their canonically conjugated momenta are Majorana-Weyl spinors. It means that each of these spinors has $16$ independent real valued components.
%cisti spinori nemaju jasnu vezu spina i statistike tako da njih ne bih stavljao u iskaz zajeno sa ovim pravim spinorima

\subsection{Choice of the background fields}

In this particular case we will work with the supermatrix $A_{MN}$ where all background fields, except RR field strength $P^{\alpha \beta}$, are constants. RR field strength will have linear coordinate dependence on bosonic coordinate $x^\mu$. With these restrictions in mind, supermatrix $A_{MN}$ has the following form

\begin{equation}
A_{MN} = \begin{bmatrix}
0 & 0 & 0 & 0 \\
0 & \kappa(\frac{1}{2}g_{\mu \nu} + B_{\mu \nu}) & \bar{\Psi}_\mu^\beta & 0\\
0 & -\Psi_\nu^\alpha & \frac{2}{\kappa}(f^{\alpha\beta} + C_\rho^{\alpha\beta} x^\rho)& 0\\
0 & 0 & 0 & 0
\end{bmatrix}.
\end{equation}  
Here $g_{\mu \nu}$ is symmetric tensor, $B_{\mu \nu}$ is Kalb-Ramond antisymmetric field, $\Psi_\mu^\alpha$ and $\bar{\Psi}_\mu^\alpha$ are Mayorana-Weyl gravitino fields, and finally, $f^{\alpha \beta}$ and $C_\rho^{\alpha \beta}$ are constants. Let us stress that dilaton field $\Phi$ is assumed to be constant, so, the factor $e^\Phi$ is included in $f^{\alpha \beta}$ and $C_\rho^{\alpha \beta}$. This will be a classical analysis and we will not calculate the dilaton shift under T-duality transformation. Based on the chirality of spinors, there are type IIA superstring theory for opposite chirality and type IIB superstring theory for same chirality.

This particular choice of supermatrix imposes following restriction on background fields

\begin{align} \label{constraint}
\gamma^\mu_{\alpha \beta} C_\mu^{\beta \gamma} = 0, \quad \gamma^\mu_{\alpha \beta} C_\mu^{\gamma \beta} = 0.
\end{align}

Remaining constraints \cite{Vertex operators} are trivial and applied only to non-physical fields. 

In addition to choice of supermatrix, in order to simplify calculation of bosonic T-duality, because all background fields are expanded in powers of $\theta^\alpha$ and $\bar\theta^\alpha$, all $\theta^{\alpha}$ and $\bar{\theta}^{\alpha}$ non-linear terms in $X^M$ and $\bar{X}^N$ will be neglected. With this in mind, components of these two vectors reduce into the following form

\begin{align}
\Pi_{\pm}^\mu \rightarrow \partial_\pm x^\mu, \quad
d_\alpha   \rightarrow  \pi_\alpha, \quad
\bar{d}_\alpha  \rightarrow  \bar{\pi} _\alpha.
\end{align}
Taking into account all these assumptions, the action (\ref{eq:dejstvo}) takes the form
\begin{equation}\label{Action full S}
\begin{gathered} 
S        =          \int_{\Sigma}   d^2   \xi          \left[     \frac{\kappa}{2}   \Pi_{ +\mu \nu } \partial_+   x^\mu  \partial_-   x^\nu          -           \pi_\alpha  ( \partial_-   \theta^\alpha   +   \Psi_\nu^\alpha   \partial_-  x^\nu )   +              (   \partial_+   \bar{ \theta }^\alpha +   \partial_+   x^\mu \bar{ \Psi }_\mu^\alpha  )   \bar{ \pi }_\alpha    \right.\\
%%%%%%%%%%%%%%%%%%%%%%%%%%%%%%%%%%%%%%%%5
\left.             +          \frac{2}{\kappa} \pi_\alpha   (   f^{ \alpha\beta }   +C_\rho^{\alpha\beta} x^\rho   ) \bar{ \pi }_\beta          \right].
\end{gathered}
\end{equation}
Here, new tensor $\Pi_{ \pm \mu \nu } = B_{\mu \nu} \pm \frac{1}{2} G_{\mu \nu}$ is introduced, where $G_{\mu \nu} = \eta_{\mu \nu} + g_{\mu \nu}$ is metric tensor. Terms for $S_{\lambda}$ and $S_{\bar{\lambda}}$ are fully decoupled from action and they will not be considered from now on.

Before considering T-duality, we can notice that fermionic momenta act as auxiliary fields in full actions. These fields can be integrated out and final action will be function of only coordinates and their derivatives. Finding equations for motion for both $\pi_\alpha$ and $\bar{ \pi }_\alpha$ we get following two equations

\begin{alignat}{2} \label{Equations for fermionic momenta 1}
\bar{ \pi }_\beta   &   =   
&&\frac{\kappa}{2}   
\left( F^{-1}  (x)   \right)_{ \beta \alpha}        
\left(       \partial_- \theta^\alpha     +    \Psi_\nu^\alpha    \partial_-   x^\nu     \right) , \\
%%%%%%%%%%%%%%%%%%%%%%%%%%%%%%%%%%%%%%%%%%%%%
\label{Equations for fermionic momenta 2}
\pi_\alpha    &   =    
-   &&\frac{\kappa}{2}
\left(     \partial_+   \bar{ \theta }^\beta     +     \partial_+   x^\mu    \bar{ \Psi }_\mu^\beta   \right)
\left( F^{-1}  (x)   \right)_{ \beta \alpha} \, ,
\end{alignat}
where $F^{\alpha\beta}(x)$ and $(F^{-1}(x))_{\alpha\beta}$ are of the form 
\begin{equation}\label{Fiinverz}
F^{\alpha\beta}(x)=f^{\alpha\beta}+C_\mu^{\alpha\beta} x^\mu\, ,\quad (F^{-1}(x))_{\alpha\beta}=( f^{ -1 } )_{ \alpha\beta }  - ( f^{ -1 } )_{ \alpha\alpha_1 }   C_\mu^{\alpha_1 \beta_1}     ( f^{ -1 } )_{ \beta_1\beta } \;x^\mu\, .
\end{equation}
In order to invert previous equations and T-dual transformation laws, as well as to simplify calculations, we take two additional assumptions. First assumption is that $C^{\alpha\beta}_\mu$ is infinitesimal. Second assumption is that $( f^{ -1 } )_{ \alpha\alpha_1 }   C_\mu^{\alpha_1 \beta_1}       ( f^{ -1 } )_{ \beta_1\beta }$ is antisymmetric under exchange of first and last index. In other words, tensor $(F^{-1}(x))_{\alpha\beta}$ has only antysimetric part that depends on $x^\mu$ and it is infinitesimal.     These assumptions are in full accordance with constraints \cite{Vertex operators}.

Substituting equations (\ref{Equations for fermionic momenta 1}) and (\ref{Equations for fermionic momenta 2}) into (\ref{Action full S}) the final form of action is

\begin{align} \label{Action final S}
S       =    
\kappa    \int_{\Sigma}    d^2    \xi      
\left[     \Pi_{ +\mu \nu }   \partial_+     x^\mu    \partial_-   x^\nu
+     \frac{1}{2}      (  \partial_+  \bar{ \theta }^\alpha  +   \partial_+   x^\mu   \bar{ \Psi }_\mu^\alpha     )
\left( F^{-1}  (x)   \right)_{ \alpha\beta }
(    \partial_-    \theta^\beta      + \Psi_\nu^\beta      \partial_-  x^\nu         )
\right] .  
\end{align}
In the following sections, this form  of action will be used for investigation of bosonic T-duality and for obtaining transformation laws between starting and T-dual coordinates.

\section{T-dualization}
\setcounter{equation}{0}

In this section T-duality will be performed along all bosonic coordinates in order to find relations that connect T-dual coordinates with coordinates and momenta of original theory. These transformation laws will then be used in subsequent chapters to find non-commutativity relations between coordinates of T-dual theory. 

Starting point for considering T-duality will be generalized Buscher T-dualization procedure \cite{Generalized Buscher procedure}. Standard Buscher procedure \cite{T-duality procedure 1,T-duality procedure 2} is designed to be applied along isometry directions on which background fields do not depend. Generalized Buscher procedure can be applied to theories with coordinate dependent background fields. The shift symmetry in the generalized procedure is localized by introduction of covariant derivatives, invariant coordinates and additional gauge fields. These newly introduced gauge fields produce additional degrees of freedom. Since we expect that starting and T-dual theory have exactly the same number of degrees of freedom we need to eliminate all excessive degrees of freedom. This is accomplished by demanding that field strength of gauge fields $(F_{+ -} = \partial_+ v_- - \partial_- v_+)$ vanishes by addition of Lagrange multipliers. Next step in procedure is fixing the gauge symmetry such that starting coordinates are constant and action is only left with gauge fields and its derivatives. From this gauge fixed action, finding equations of motion for gauge fields, expressing gauge fields as function of Lagrange multipliers and inserting those equations into action we can obtain T-dual action, were Lagrange multipliers of original theory now play the role of T-dual coordinates.  

In cases where shift symmetry is absent, T-duality can still be performed by extending generalized Buscher procedure \cite{Auxiliary action Buscher procedure}. This extension is based on replacing original action with translation invariant auxiliary action. Form of this auxiliary action is exactly the same as the form of action where translation symmetry was localized and gauged fixed, that is, derivatives have been replaced with gauge fields and coordinates with integrals of gauge fields. Auxiliary action gives correct T-dual theory only if original action can be salvaged from it. In cases where this is possible, original theory is obtained by finding equations of motion with respect to Lagrange multipliers and inserting their solutions into auxiliary action.

Action (\ref{Action final S}) is invariant to translation symmetry, by the virtue of antisymmetric part of $F^{-1}_{\alpha \beta}$, tensor $(f^{-1}C_\mu f^{-1})_{\alpha\beta}$. Following antisymmetricity of this tensor, we can rewrite the action (\ref{Action final S}) in the following way
\begin{equation}\label{Actionedd}
S       =    
\kappa    \int_{\Sigma}    d^2    \xi      
\left[     \Pi_{ +\mu \nu }   \partial_+     x^\mu    \partial_-   x^\nu
+     \frac{1}{2}      \epsilon^{mn}\partial_m(  \bar{ \theta }^\alpha  +     x^\mu   \bar{ \Psi }_\mu^\alpha     )
\left( F^{-1}  (x)   \right)_{ \alpha\beta }
\partial_n(        \theta^\beta      + \Psi_\nu^\beta        x^\nu         )
\right] .
\end{equation}

Let us now consider the global shift symmetry $\delta x^\mu=\lambda^\mu$ and vary the action (\ref{Actionedd})
\begin{equation}
\delta S=-\frac{\kappa}{2}(f^{-1}C_\mu f^{-1})_{\alpha\beta}\lambda^\mu \int_\Sigma d^2\xi \epsilon^{mn}\partial_m (\bar\theta^\alpha+\bar\Psi^\alpha_\nu x^\nu)\partial_n (\theta^\beta+\Psi^\beta_\rho x^\rho)\, ,
\end{equation}
where $m,n$ are indices of the twodimensional worldsheet. After one partial integration, we obtain one surface term and one term which is identically zero because it is summation of symmetric, $\partial_m \partial_n$, and antisymmetric, $\epsilon^{mn}$, tensor. The surface term is zero for trivial topology. So, the shift isometry exists.

In order to find T-dual action we have to implement following substitutions

\begin{alignat}{2}
&\partial_\pm x^\mu     &&\rightarrow       D_\pm x^\mu=\partial_\pm x^\mu+v_\pm^\mu, \\ 
%%%%%%%%%%%%%%%%%%%%%%%%%%%%%%%%%%%%%%%%%%%%%%%%%%%%%%%%%
&x^\rho                  &&\rightarrow       x^\mu_{inv}=\int_P d\xi^m D_m x^\mu=x^\mu(\xi)-x^\mu(\xi_0)+\Delta V^\mu \, ,    \Delta V^\mu=\int_{P} d { \xi }^m v_m^\rho (\xi)\label{V expresion},\\ 
%%%%%%%%%%%%%%%%%%%%%%%%%%%%%%%%%%%%%%%%%%%%%%%%%%%%%%%
&S                    &&\rightarrow     S  +  \frac{\kappa}{2}    \int_{\Sigma}       d^2       \xi          \left[ v_+^\mu \partial_- y_\mu   -  v_-^\mu   \partial_+ y_\mu
\right].
\end{alignat}
Because of the shift symmetry we fix the gauge, $x^\mu(\xi)=x^\mu(\xi_0)$ and, inserting these substitutions into action (\ref{Action final S}), we obtain auxiliary action suitable for T-dualization
\begin{equation}\label{Auxiliary action S_{aux}}
\begin{gathered}  
S_{aux}       =    \kappa    \int_{\Sigma}    d^2    \xi
\left[ \Pi_{ +\mu \nu } v_+^\mu v_-^\nu    +    \frac{1}{2} (      \partial_+     \bar{ \theta }^\alpha      +     v_+^\mu    \bar{ \Psi }_\mu^\alpha     )         \left( F^{-1}  ( \Delta V )   \right)_{ \alpha\beta }   (      \partial_-\theta^\beta     + \Psi_\nu^\beta  v_-^\nu       ) \right. \\
%%%%%%%%%%%%%%%%%%%%%%%%%%%%%%%%%%%%%%%%%%%%%%%%%%%%%%%
\left.  +      \frac{1}{2}(     v_+^\mu    \partial_- y_\mu           -         v_-^\mu       \partial_+      y_\mu       )
\right].
\end{gathered}
\end{equation}
It should be noted that, path $P$ that is taken in expression for $\Delta V^\rho$ goes from some starting point $\xi_0$ to end point $\xi$. Introduction of this element makes this action non-local, however, this is a necessary step in order to find T-dual theory of coordinate dependent background fields \cite{Generalized Buscher procedure}. 

In order to check if substitutions we had introduced are valid and that they will lead to correct T-dual theory of starting action, we need to be able to obtain original action by finding solutions to equations of motion for Lagrange multipliers.
Equations of motion for Lagrange multipliers give us

\begin{equation} \label{x form v}
\partial_- v_+^\mu - \partial_+ v_-^\mu = 0\quad \Rightarrow \quad v_\pm^\mu = \partial_\pm x^\mu.
\end{equation}
Inserting this result into (\ref{V expresion}) we get the following
\begin{equation} \label{X from V}
\Delta V^\rho = \int_{P} d { \xi^\prime }^m \partial_m x^\rho (\xi^\prime) = x^\rho (\xi) - x^\rho (\xi_0) = \Delta x^\rho.
\end{equation}
Since, we had shift symmetry in original action, we can let $x^\rho (\xi_0)$ be any arbitrary constant. Taking all this into account and inserting (\ref{x form v}), (\ref{X from V}) into (\ref{Auxiliary action S_{aux}}) we obtain our starting action (\ref{Action final S}).

Before we obtain equations for motion for gauge fields, we would like to make following substitution in action

\begin{gather}
Y_{+ \mu} = \partial_+ y_\mu - \partial_+ \bar{ \theta }^\alpha \left( F^{-1}  ( \Delta V )   \right)_{ \alpha\beta } \Psi^\beta_\mu, \qquad Y_{-\mu} = \partial_- y_\mu +\bar{ \Psi }^\alpha_\mu \left( F^{-1}  ( \Delta V )   \right)_{ \alpha\beta } \partial_- \theta^\beta,\\
\bar{\Pi}_{+\mu\nu} = \Pi_{ +\mu \nu } + \frac{1}{2} \bar{ \Psi }_\mu^\alpha \left( F^{-1}  ( \Delta V )   \right)_{ \alpha\beta }\Psi_\nu^\beta =  \breve{\Pi}_{+ \mu\nu} - \frac{1}{2}\bar{ \Psi }_\mu^\alpha
( f^{ -1 } )_{ \alpha\alpha_1 }   C_\rho^{\alpha_1 \beta_1}     ( f^{ -1 } )_{ \beta_1\beta }
 \Psi_\nu^\beta \Delta V^\rho,\\
 \breve{\Pi}_{+ \mu\nu} \equiv \Pi_{ +\mu \nu } + \frac{1}{2} \bar{ \Psi }_\mu^\alpha  (f^{-1})_{ \alpha\beta }\Psi_\nu^\beta.
\end{gather}

With these substitutions in mind we have that auxiliary action takes the following form

\begin{equation}\label{Auxiliary action S_{aux} 2}
\begin{gathered}  
S_{aux}       =    \kappa    \int_{\Sigma}    d^2    \xi
\left[ \bar{ \Pi }_{ +\mu \nu } v_+^\mu v_-^\nu         +\frac{1}{2}v_+^\mu Y_{- \mu} - \frac{1}{2}v_-^\mu Y_{+\mu}     +    \frac{1}{2}      \partial_+     \bar{ \theta }^\alpha    \left( F^{-1}  ( \Delta V )   \right)_{ \alpha\beta }        \partial_-\theta^\beta
\right].
\end{gathered}
\end{equation}

This action produces following equations of motion for gauge fields

\begin{equation}\label{equations of motion of v}
\bar{ \Pi }_{+\mu\nu} v_-^\nu = -( \frac{1}{2}Y_{-\mu} + \beta^+_{\mu}(V) ), \qquad \bar{ \Pi }_{+\mu \nu}v_+^\mu = \frac{1}{2}Y_{+\nu} -\beta^-_{\nu}(V) .
\end{equation}

Here, function $\beta^\pm(V)$ is obtained from variation of term containing $\Delta V^\rho$ in expression for $ F^{-1}  ( \Delta V )$ (details are presented in Appendix \ref{appendix A})

\begin{align}\label{beta 1}
\beta^-_{\mu}(V) = & \frac{1}{4}
\partial_+\big[   \bar{ \theta }^\alpha      +     V^{\nu_1}     \bar{ \Psi }_{\nu_1} ^\alpha    \big]  
( f^{ -1 } )_{ \alpha\alpha_1 }     C_\mu^{\alpha_1 \beta_1}      ( f^{ -1 } )_{ \beta_1\beta }
\big[    \theta^\beta    + \Psi_{\nu_2} ^\beta  V^{\nu_2}        \big]\nonumber
\\
-&\frac{1}{4}
\big[ \bar{ \theta }^\alpha     +     V^{\nu_1}     \bar{ \Psi }_{\nu_1} ^\alpha    \big]  
( f^{ -1 } )_{ \alpha\alpha_1 }     C_\mu^{\alpha_1 \beta_1}      ( f^{ -1 } )_{ \beta_1\beta } 
\partial_+\big[  \theta^\beta   + \Psi_{\nu_2} ^\beta  V^{\nu_2}       \big],\\
%%%%%%%%%%%%%%%%%%%%%%%%%%%%%%%%%%%%%%%%%%%%%%%%%%%%%%%%%%%%%%%%
\beta^+_{\mu}(V) \label{beta 2}
=&  \frac{1}{4}\big[ \bar{ \theta }^\alpha     +     V^{\nu_1}     \bar{ \Psi }_{\nu_1} ^\alpha    \big]  
( f^{ -1 } )_{ \alpha\alpha_1 }     C_\mu^{\alpha_1 \beta_1}      ( f^{ -1 } )_{ \beta_1\beta } 
\partial_-\big[  \theta^\beta   + \Psi_{\nu_2} ^\beta  V^{\nu_2}       \big]\nonumber
\\
-&\frac{1}{4}
\partial_-\big[   \bar{ \theta }^\alpha      +     V^{\nu_1}     \bar{ \Psi }_{\nu_1} ^\alpha    \big]  
( f^{ -1 } )_{ \alpha\alpha_1 }     C_\mu^{\alpha_1 \beta_1}      ( f^{ -1 } )_{ \beta_1\beta }
\big[    \theta^\beta    + \Psi_{\nu_2} ^\beta  V^{\nu_2}        \big].
\end{align}
Here we have took advantage of the fact that $\partial_\pm V^\mu = v_\pm^\mu$ (more details in Appendix \ref{appendix A}).  Let us note that $V^\mu$ in the expressions for beta functions is actually $V^{(0)\mu}$ because it stands besides $C^{\alpha\beta}_\mu$. We omit index $(0)$ just in order to simplify the form of the expressions.

In order to find how gauge fields depend on Lagrange multipliers, we need to invert equations of motion (\ref{equations of motion of v}). Since $C_\mu^{\alpha \beta}$ is an infinitesimal constant, these equations can be inverted iteratively \cite{Inversion of equations of motion}. We separate variables into two parts, one finite and one proportional to $C_\mu^{\alpha \beta}$. After doing this we have

\begin{equation}\label{v as function of y}
v_-^\nu = - \bar{\Theta}^{\nu\mu}_-\Big[\frac{1}{2}Y_{-\mu} + \beta^+_{\mu}(V^{(0)})\Big], \qquad v_+^\mu = \Big[\frac{1}{2}Y_{+\nu} -\beta^-_{\nu}(V^{(0)})\Big]\bar{\Theta}^{\nu\mu}_-.
\end{equation}

Functions $\beta_{\pm\mu}(V^{(0)})$ are obtained by substituting first order of expression for $v_\pm$ into $\beta_{\pm\mu}(V)$, where $V^{(0)}$ is given by

\begin{align}
\begin{gathered}
\Delta V^{(0) \rho}  = \int_{P} d \xi^m v_m^{(0)\rho} \\   
%%%%%%%%%%%%%%%%%%%%%%%%%%%%%%%%%%%%%%%%%%%%%%%%%%%%%%%%%%%%%%%%
=\frac{1}{2} \int_{P} d \xi^+ \breve{\Theta}_-^{\rho_1 \rho} \left[ \partial_+ y_{\rho_1}     -  \partial_+ \bar{ \theta } ^\alpha ( f^{- 1 } )_{ \alpha\beta } \Psi_{\rho_1}^\beta
\right]
%%%%%%%%%%%%%%%%%%%%%%%%%%%%%%%%%%%%%%%%%%%%%%%%%%%%%%%%%%%%%%%
- \frac{1}{2} \int_{P} d \xi^- \breve{\Theta}_-^{\rho \rho_1} 
\left[    \partial_- y_{\rho_1}  +  \bar{ \Psi }_{\rho_1}^\alpha ( f^{- 1 } )_{ \alpha\beta } \partial_- \theta^\beta
\right] .
\end{gathered}
\end{align}

Where $\bar{\Theta}_-^{\mu \nu}$ is inverse tensor of $\bar{\Pi}_{+\mu\nu} = \Pi_{ +\mu \nu } + \frac{1}{2} \bar{ \Psi }_\mu^\alpha \left( F^{-1}  ( \Delta V )   \right)_{ \alpha\beta }\Psi_\nu^\beta $, defined as

\begin{equation}
\bar{\Theta}_-^{\mu\nu} \bar{\Pi}_{+\nu\rho} = \delta^\mu_\rho,
\end{equation}
where
\begin{gather}
\bar{\Theta}_-^{\mu\nu} = \breve{\Theta}_-^{\mu \nu}  + \frac{1}{2} \breve{\Theta}_-^{\mu \mu_1} \bar{ \Psi }_{\mu_1}^\alpha (f^{-1})_{\alpha \alpha_1} C^{\alpha_1 \beta_1}_\rho V^{(0)\rho} (f^{-1})_{\beta_1\beta} \Psi_{\nu_1}^{\beta_1}  \breve{\Theta}_-^{\nu_1 \nu} ,\\
%%%%%%%%%%%%%%%%%%%%%%%%%%%%%%%%%%%%%%%%%%%%%%%%%%%%%%%%%%%%%%%%%%%%%%%%%%%%%%%%%%%%%%%%%%%%%%%%%%%%%%%%%%%%%%%%%%%%%%%%%%%%%%%%%
\breve{\Theta}_-^{\mu \nu} \breve{\Pi}_{\nu \rho} = \delta^\mu_\rho,\qquad \breve{\Theta}_-^{\mu\nu} = \Theta_-^{\mu\nu} - \frac{1}{2} \Theta_-^{\mu\mu_1} \bar{ \Psi }_{\mu_1}^\alpha (\bar{f}^{-1})_{\alpha\beta} \Psi^\beta_{\nu_1} \Theta_-^{\nu_1\nu}\\
%%%%%%%%%%%%%%%%%%%%%%%%%%%%%%%%%%%%%%%%%%%%%%%%%%%%%%%%%%%%%%%%%%%%%%%%%%%%%%%%%%%%%%%%%%%%%%%%%%%%%%%%%%%%%%%%%%%%%%%%%%%%%%%%%
\bar{f}^{\alpha\beta} = f^{\alpha \beta} + \frac{1}{2} \Psi_\mu^\alpha \Theta_-^{\mu\nu} \bar{ \Psi }_\nu^\beta,\\
%%%%%%%%%%%%%%%%%%%%%%%%%%%%%%%%%%%%%%%%%%%%%%%%%%%%%%%%%%%%%%%%%%%%%%%%%%%%%%%%%%%%%%%%%%%%%%%%%%%%%%%%%%%%%%%%%%%%%%%%%%%%%%%%%
\Theta_-^{\mu\nu} \Pi_{ +\mu \rho } = \delta^\mu_\rho, \qquad \Theta_- = -4 (G_E^{-1} \Pi_- G^{-1})^{\mu\nu}\, .
\end{gather}
Tensor $G_{E \mu\nu} \equiv G_{\mu \nu} -4(BG^{-1}B)_{\mu\nu} $ is known in the literature as the effective metric.

Inserting equations (\ref{v as function of y}) into (\ref{Auxiliary action S_{aux}}), keeping only terms that are linear in $C_\mu^{\alpha \beta}$ we obtain T-dual action

\begin{equation}
\label{S_{T-dual}}
S^\star = \frac{\kappa}{2}\int_{\Sigma} \Big[ \frac{1}{2}\bar{\Theta}_-^{\mu \nu} Y_{+ \mu} Y_{- \nu} + \partial_+ \bar{ \theta }^\alpha 
\left( F^{-1}  ( \Delta V )   \right)_{ \alpha\beta } \partial_- \theta^\beta   \Big] .
\end{equation}
Comparing starting action (\ref{Action final S}) with T-dual action, were we note that $\partial_\pm x^\mu$ transforms into $\partial_\pm y_\mu$ and $x^\mu$ transforms into $V^{(0)}$, we can deduce that T-dual action has following arguments.

\begin{gather}
{}^\star\bar{ \Pi }_+^{\mu \nu} = \frac{1}{4} \bar{\Theta}_-^{\mu \nu},
\\
\left( {}^\star F^{-1} \text{\small $( V^{(0)} )$}    \right)_{ \alpha\beta } = \left( F^{-1}  \text{\small $( V^{(0)} )$}   \right)_{ \alpha\beta } - \frac{1}{2} \left( F^{-1} \text{\small $( V^{(0)} )$}   \right)_{ \alpha\alpha_1 } \Psi_\mu^{\alpha_1} \bar{\Theta}_-^{\mu \nu} \bar{ \Psi }_\nu^{\beta_1} \left( F^{-1}  \text{\small $( V^{(0)} )$}    \right)_{ \beta_1\beta },
\\%%%%%%%%%%%%%%%%%%%%%%%%%%%%%%%%%%%%%%%%%%%%%%%%%%%%%%%%%%%
{}^\star \bar{ \Psi }^{\mu \alpha}  \left( {}^\star F^{-1}  \text{\small $( V^{(0)} )$}    \right)_{ \alpha\beta } = \frac{1}{2} \bar{\Theta}_-^{\mu \nu} \bar{ \Psi }_\nu^\alpha \left( F^{-1}  \text{\small $( V^{(0)} )$}    \right)_{ \alpha\beta },\\
\left( {}^\star F^{-1}  \text{\small $( V^{(0)} )$}    \right)_{ \alpha\beta } {}^\star \Psi^{\nu \beta} = - \frac{1}{2} \left( F^{-1}  \text{\small $( V^{(0)} )$}    \right)_{ \alpha\beta } \Psi_\mu^\beta \bar{\Theta}_-^{\mu \nu}.
\end{gather}

In order to express T-dual gravitino background fields in terms of its components, it is useful to calculate inverse of field $ {}^\star F^{-1}_{ \alpha\beta }$

\begin{equation}
{}^\star F^{\alpha\beta}\text{\small $( V^{(0)} )$} = F^{\alpha \beta} \text{\small $( V^{(0)} )$}  + \frac{1}{2} \Psi_\mu^\alpha \Theta_-^{\mu \nu} \bar{ \Psi }_\nu^\beta .
\end{equation}
With this equation at hand it is straightforward to obtain T-dual gravitino fields. Here we present T-dual gravitino fields expanded in terms of their components 
\begin{align}
\label{Background fields transformation 2}
%&\begin{gathered}
% {}^\star F^{\alpha\beta}  \text{\small $( V^{(0)} )$}  %=F^{\alpha\beta} \text{\small $( V^{(0)} )$}   - \frac{1}{2}    %C_\rho^{\alpha \alpha_1} \text{\small $ V^{(0)\rho} $} ( f^{- 1 %} )_{ \alpha_1\alpha_2}   \Psi_\mu^{ \alpha_2 }  %\bar{\Theta}_-^{\mu \nu} \bar{ \Psi }_{\nu}^{\beta}  \\
%\begin{aligned}
%&+ \frac{1}{2}   \Psi_\mu^{ \alpha}  \bar{\Theta}_-^{\mu \nu} %\bar{ \Psi }_{\nu}^{\beta}  - \frac{1}{2}   \Psi_\mu^{ \alpha %}  \bar{\Theta}_-^{\mu \nu} \bar{ \Psi }_{\nu}^{\beta_2} ( f^{- %1 } )_{ \beta_2\beta_1 }   C_\rho^{\beta_1 \beta} \text{\small %$ V^{(0)\rho} $} ,
%%%%%%%%%%%%%%%%%%%%%%%%%%%%%%%%%%%%%%%%%%%%%%%%%%%%%%%%%%%%%%%%%%%%%%%%%%%%%%%%%%%%%%%%%%%%%%%%%%%%%%%%%%%%%%%%%%%%%%%%%%%%%%%%%%%%%%%%%%%%%%%%%%%%%%%%%%%%%%%%%%%%%%%%%%%%%%%%%%%%%%%%%%%%%%%%%%
%\end{aligned}
%\end{gathered}\\
&\begin{gathered}
\begin{aligned}
\quad{}^\star \bar{ \Psi }^{\mu \alpha} = \frac{1}{2} \bar{\Theta}^{\mu\nu} \bar{ \Psi }_\nu^\alpha + \frac{1}{4} \bar{\Theta}_-^{\mu \mu_1} \bar{ \Psi }_{\mu_1}^\beta \left( F^{-1}  \text{\small $( V^{(0)} )$}    \right)_{ \beta\beta_1 }
\Psi_\nu^{\beta_1} \Theta_-^{\nu \nu_1} \bar{ \Psi }_{\nu_1}^\alpha,
\end{aligned}
\end{gathered}\\
&\begin{gathered}
\begin{aligned}
\quad{}^\star \Psi^{\nu \beta} = -\frac{1}{2} \Psi_\mu^\beta \bar{\Theta}_-^{\mu\nu} -\frac{1}{2} \Psi_\mu^\beta \Theta_-^{\mu \mu_1} \bar{ \Psi }_{\mu_1}^{\alpha}  \left( F^{-1}  \text{\small $( V^{(0)} )$}   \right)_{ \alpha\alpha_1 }\Psi_{\nu_1}^{\alpha_1} \bar{\Theta}_-^{\nu_1 \nu}.
\end{aligned}
\end{gathered}
\end{align}   
The general conclusion is that all background fields get the linear corrections in $C^{\alpha\beta}_\mu$ comparing with the results of the case with constant background fields \cite{nasareferenca}. Also the coordinate dependence is present in all T-dual background fields.

From the above equations we see how background fields of original theory transform under T-duality. It should be noted that these actions are of the same form taking into account that initial coordinates $x^\mu$ are replaced by $y_\mu$ after T-dualization.

\section{T-dualization of T-dual theory}
\setcounter{equation}{0}

From requirement that original theory and T-dual theory be physically equivalent, it should be possible to obtain original theory from T-dual one by applying T-duality procedure a second time. Since original action possessed translation symmetry, we have that this symmetry is inherited by T-dual action. T-dual theory is invariant to translations of T-dual coordinate. However, even in cases where starting action is not invariant to translation symmetry we can expect emergence of this symmetry in T-dual theory. This is a natural consequence of introducing $\Delta V^{(0)}$ and of the fact that T-dual theory is intrinsically a non-local one. T-dualization of T-dual theory is obtained with generalized Buscher procedure and steps are identical as before.

\begin{align}
\partial_\pm y_\mu       \rightarrow       &   D_\pm y_\mu  = \partial_\pm y_\mu +  u_{\pm \mu}  \rightarrow D_\pm y_\mu =  u_{\pm \mu}, \\
%%%%%%%%%%%%%%%%%%%%%%%%%%%%%%%%%%%%%%%%%%%%%%%%%%%%%%%%%%%%%%%%
\Delta V^{(0)\rho}       \rightarrow     &   \Delta U^{(0) \rho}, \\
%%%%%%%%%%%%%%%%%%%%%%%%%%%%%%%%%%%%%%%%%%%%%%%%%%%%%%%%%%%%%%%
\Delta U^{ (0)\rho}  = 
&\frac{1}{2} \int_{P} d \xi^+ \breve{\Theta}_-^{\rho_1 \rho} \left[ u_{+ \rho_1}     -  \partial_+ \bar{ \theta } ^\alpha ( f^{- 1 } )_{ \alpha\beta } \Psi_{\rho_1}^\beta
\right] \nonumber \\
- &\frac{1}{2} \int_{P} d \xi^- \breve{\Theta}_-^{\rho \rho_1} 
\left[    u_{- \rho_1}  +  \bar{ \Psi }_{\rho_1}^\alpha ( f^{- 1 } )_{ \alpha\beta } \partial_- \theta^\beta 
\right]\, ,\\ 
Y_{+ \mu} \rightarrow &U_{+ \mu} = u_{+ \mu} - \partial_+ \bar{ \theta }^\alpha \left( F^{-1}  \text{\small $( \Delta U^{(0)} )$}    \right)_{ \alpha\beta } \Psi_\mu^\beta\\
Y_{- \mu} \rightarrow &U_{- \mu} = u_{- \mu} + \bar{ \Psi }_\mu^\alpha \left( F^{-1}  \text{\small $( \Delta U^{(0)} )$}    \right)_{ \alpha\beta } \partial_- \theta^\beta\\
{}^\star S \rightarrow & {}^\star S + \frac{\kappa}{2} \int_\Sigma d^2\xi  (u_{+ \mu}  \partial_-x^\mu  - u_{- \mu}\partial_+ x^\mu  ) \, .
\end{align}

In first line we immediately fixed gauge by choosing $y(\xi) = const$. Inserting these substitutions into (\ref{S_{T-dual}}) we get

\begin{equation}
\label{S_{gauge fix}}
{}^\star S_{gfix} =\frac{k}{2}\int_{\Sigma} d^2 \xi\Big[ \frac{1}{2}\bar{\Theta}_-^{\mu \nu} U_{+ \mu} U_{- \nu} + \partial_+ \bar{ \theta }^\alpha 
\left( F^{-1}  \text{\small $( \Delta U^{(0)} )$}   \right)_{ \alpha\beta } \partial_- \theta^\beta  + (u_{+ \mu}  \partial_-x^\mu  - u_{- \mu}\partial_+ x^\mu  ) \Big] .
\end{equation}
Finding equations of motion for Lagrange multipliers and inserting solution to those equations into gauge fixed action we return to the starting point of this chapter, T-dual action. On the other hand, finding equations of motion for gauge fields

\begin{align}\label{T^2 equations of motion 1}
u_{+ \mu} &= 2 \Big[ \partial_+ x^\nu \bar{ \Pi }_{+ \nu \mu}  + \beta^-_{\mu}(x) 
\Big] + \partial_+ \bar{ \theta }^\alpha  \left(F^{-1} ( x)\right)_{\alpha \beta} \Psi_\mu^\beta, \\
\label{T^2 equations of motion 2}
u_{- \nu} &= - 2 \Big[ \bar{ \Pi }_{+ \mu \nu} \partial_-x^\nu +\beta^+_{\mu}(x)
\Big]- \bar{ \Psi }_\nu^\alpha \left(F^{-1} (x)\right)_{\alpha \beta} \partial_- \theta^\beta \, ,
\end{align}
and inserting these equations into the gauge fixed action, keeping all terms linear with respect to $C_\rho^{\mu\nu}$,, we obtain our original action (\ref{Action final S}). Here we use the freedom to choose $\Delta x^\mu = x(\xi) - x(\xi_0)$, with  $x(\xi_0) = 0$.

\section{Non-commutative relations}
\setcounter{equation}{0}

Having found T-dual action and equations that link T-dual coordinate with original coordinates in previous chapters, in this chapter we will focus on establishing a relationship between Poisson brackets of original and T-dual theory. Furthermore, we will mainly focus on Poisson brackets between bosonic variables and their momenta. Original theory is a geometric one with variables $x^\mu(\xi)$ and $\pi_\mu(\xi)$. Therefore, it is natural to impose standard Poisson structure on original theory

\begin{equation}
\{ x^\mu (\sigma), \pi_\nu (\bar{\sigma})  \} = \delta^\mu_\nu \delta(\sigma - \bar{\sigma}), \quad \{ x^\mu(\sigma) , x^\nu(\bar{\sigma}) \} = 0, \quad \{ \pi_\mu(\sigma), \pi_\nu(\bar{\sigma})  \} = 0.
\end{equation}

In order to find Poisson brackets of T-dual theory, we need to find T-dual transformation laws which connect the initial and T-dual coordinates.  Starting with relations (\ref{T^2 equations of motion 1}) and (\ref{T^2 equations of motion 2}) and using equations of motion for Lagrange multipliers $x^\mu$, $u_{\pm \mu}=\partial_\pm y_\mu$, we obtain T-dual transformation laws
\begin{equation}
\partial_+ y_\mu \cong 2 \Big[ \partial_+ x^\nu \bar{ \Pi }_{+ \nu \mu}  + \beta^-_{\mu}(x) 
\Big] + \partial_+ \bar{ \theta }^\alpha  \left(F^{-1} ( x)\right)_{\alpha \beta} \Psi_\mu^\beta\, ,
\end{equation}
\begin{equation}
\partial_- y_\mu \cong - 2 \Big[ \bar{ \Pi }_{+ \mu \nu} \partial_-x^\nu +\beta^+_{\mu}(x)
\Big]- \bar{ \Psi }_\nu^\alpha \left(F^{-1} (x)\right)_{\alpha \beta} \partial_- \theta^\beta\, ,
\end{equation}
where symbol $\cong$ denotes T-dual transformation. Subtracting these two equations, we get
\begin{equation}
y'_\mu\cong \bar\Pi_{+\mu\nu}\partial_- x^\nu + \partial_+ x^\nu \bar\Pi_{+\nu\mu}+\beta^+_\mu+\beta^-_\mu+\frac{1}{2}\partial_+ \bar{ \theta }^\alpha  \left(F^{-1} ( x)\right)_{\alpha \beta} \Psi_\mu^\beta+\frac{1}{2}\bar{ \Psi }_\nu^\alpha \left(F^{-1} (x)\right)_{\alpha \beta} \partial_- \theta^\beta\, .
\end{equation}
Taking into account that bosonic momenta, $\pi_\mu$ of original theory are of the form
\begin{equation}
\pi_\mu = \kappa\Big[ \bar\Pi_{+\mu\nu}\partial_- x^\nu + \partial_+ x^\nu \bar\Pi_{+\nu\mu} + \frac{1}{2} \bar{ \Psi }_\mu^\alpha \left( F^{-1}  (x)    \right)_{ \alpha\beta } \partial_- \theta^\beta + \frac{1}{2}\partial_+ \bar{ \theta }^\alpha \left( F^{-1}  (x)    \right)_{ \alpha\beta }\Psi_{\nu}^\beta \Big],
\end{equation}
and $\beta^0_\mu=\beta^+_\mu+\beta^-_\mu$, we obtain
\begin{equation}\label{sigma derivative of y}
y'_\mu \cong\frac{\pi_\mu}{\kappa} +\beta^0_\mu(x)\, .
\end{equation}
Here $\beta^0_\mu(x)$ is given by

\begin{align}
\beta^0_{\mu}(x) = & \frac{1}{2}
\partial_\sigma\big[   \bar{ \theta }^\alpha      +     x^{\nu_1}     \bar{ \Psi }_{\nu_1} ^\alpha    \big]  
( f^{ -1 } )_{ \alpha\alpha_1 }     C_\mu^{\alpha_1 \beta_1}      ( f^{ -1 } )_{ \beta_1\beta }
\big[    \theta^\beta    + \Psi_{\nu_2} ^\beta  x^{\nu_2}        \big] \nonumber
\\
-&\frac{1}{2}
\big[ \bar{ \theta }^\alpha     +     x^{\nu_1}     \bar{ \Psi }_{\nu_1} ^\alpha    \big]  
( f^{ -1 } )_{ \alpha\alpha_1 }     C_\mu^{\alpha_1 \beta_1}      ( f^{ -1 } )_{ \beta_1\beta } 
\partial_\sigma\big[  \theta^\beta   + \Psi_{\nu_2} ^\beta  x^{\nu_2}       \big]\, .
\end{align}

To find Poisson bracket between T-dual coordinates, we can start by finding Poisson bracket of sigma derivatives of T-dual coordinates and then integrating twice (see \cite{Generalized Buscher procedure 3}, \cite{Integration of Poisson brackets}, \cite{Integration of Poisson brackets 2}, \ref{appendix B} ). Implementing this procedure we have that Poisson bracket is given as

\begin{align}\label{Poisson brackets T-dual}
\begin{gathered}
\{ y_{\nu_1} (\sigma),  y_{\nu_2} (\bar{\sigma})  \} \cong\\
\begin{aligned}
\frac{1}{2k}[2 \delta^{\mu_1}_{\nu_1}\delta^{\mu_2}_{\nu_2} - \delta^{\mu_2}_{\nu_1}\delta^{\mu_1}_{\nu_2}]  \Big[ K_{\mu_1 \mu_2} (\bar{\sigma}) + K_{\mu_2 \mu_1} (\sigma)  \Big]H(\sigma - \bar{\sigma}) \, ,
\end{aligned}
\end{gathered}
\end{align}
where, for the sake of simplicity, we introduced

\begin{align}
\begin{gathered} \label{Shortcut}
K_{\mu \nu} (\sigma) = 
\Big(\bar{ \theta }^\alpha(\sigma)  +x^{\mu_1}(\sigma)  \bar{ \Psi }_{\mu_1}^\alpha
\Big) 
( f^{ -1 } )_{ \alpha\alpha_1 }     C_\mu^{\alpha_1 \beta_1}      ( f^{ -1 } )_{ \beta_1\beta } \Psi_{\nu}^\beta\\
-\bar{ \Psi }_{\nu}^\alpha
( f^{ -1 } )_{ \alpha\alpha_1 }     C_\mu^{\alpha_1 \beta_1}      ( f^{ -1 } )_{ \beta_1\beta } 
\Big(
\theta^\beta (\sigma) + \Psi_{\nu_1}^\beta  x^{\nu_1} (\sigma)
\Big).
\end{gathered}
\end{align}
%\begin{align} \label{Poisson brackets T-dual}
%\begin{gathered}
%\{ y_{\nu_1} (\sigma),  y_{\nu_2} (\bar{\sigma})  \} \cong  \\
%%%%%%%%%%%%%%%%%%%%%%%%%%%%%%%%%%%%%%%%%%%%%%%%%%%%%%%%%%%%%%%%%
%\begin{aligned}
%%%%%%%%%%%%%%%%%%%%%%%%%%%%%%%%%%%%%%%%%%%%%%%%%%%%%%%%%%%%%%%%%
%& \frac{2}{k} \big( \delta_{\nu_1}^\rho \delta_{\nu_2}^\gamma + %\delta_{\nu_2}^\rho \delta_{\nu_1}^\gamma \big)  \bar{ \Psi %}_{\rho}^\alpha
%( f^{ -1 } )_{ \alpha\alpha_1 }     C_\gamma^{\alpha_1 %\beta_1}      ( f^{ -1 } )_{ \beta_1\beta } 
%\Big(
%\theta^\beta (\sigma) + \Psi_{\nu}^\beta  x^{\nu} (\sigma)
%\Big) H(\sigma - \bar{\sigma}) \\
%%%%%%%%%%%%%%%%%%%%%%%%%%%%%%%%%%%%%%%%%%%%%%%%%%%%%%%%%%%%%%%%
%-& \frac{2}{k} \Big(
%\bar{ \theta }^\alpha(\sigma)  +x^{\mu}(\sigma)  \bar{ \Psi %}_{\mu}^\alpha
%\Big) 
%( f^{ -1 } )_{ \alpha\alpha_1 }     C_\gamma^{\alpha_1 %\beta_1}      ( f^{ -1 } )_{ \beta_1\beta } \Psi_{\rho_1}^\beta %\big( \delta_{\nu_1}^\rho \delta_{\nu_2}^\gamma + %\delta_{\nu_2}^\rho \delta_{\nu_1}^\gamma \big) H(\sigma - %\bar{\sigma})\, .
%\end{aligned}
%\end{gathered}
%\end{align}
Here, $H(\sigma - \bar{\sigma})$ is same step function defined in Appendix \ref{appendix B}. It should be noted that these Poisson brackets are zero when $\sigma = \bar{\sigma}$. However, in cases where string in curled around compactified dimension, that is cases where $\sigma - \bar{\sigma} = 2\pi$, we have following situation

\begin{align}
\begin{gathered}
\{ y_{\nu_1} (\sigma + 2\pi),  y_{\nu_2} (\sigma)  \} \cong \frac{1}{2k}[2 \delta^{\mu_1}_{\nu_1}\delta^{\mu_2}_{\nu_2} - \delta^{\mu_2}_{\nu_1}\delta^{\mu_1}_{\nu_2}]  \Big[ K_{\mu_1 \mu_2} (\sigma) + K_{\mu_2 \mu_1} (\sigma)  \Big] \\
%%%%%%%%%%%%%%%%%%%%%%%%%%%%%%%%%%%%%%%%%%%%%%%%%%%%%%%%%%%%%%%%%
\begin{aligned}
%%%%%%%%%%%%%%%%%%%%%%%%%%%%%%%%%%%%%%%%%%%%%%%%%%%%%%%%%%%%%%%%%
%&\frac{2}{k} \big( \delta_{\nu_1}^\rho \delta_{\nu_2}^\gamma + %\delta_{\nu_2}^\rho \delta_{\nu_1}^\gamma \big)  \bar{ \Psi %}_{\rho}^\alpha
%( f^{ -1 } )_{ \alpha\alpha_1 }     C_\gamma^{\alpha_1 %\beta_1}      ( f^{ -1 } )_{ \beta_1\beta } 
%\Big(
%\theta^\beta (\sigma) + \Psi_{\nu}^\beta  x^{\nu} (\sigma) + %2\pi\Psi_{\nu}^\beta  N^\nu
%\Big)\\
%%%%%%%%%%%%%%%%%%%%%%%%%%%%%%%%%%%%%%%%%%%%%%%%%%%%%%%%%%%%%%%%%
%-& \frac{2}{k} \Big(
%\bar{ \theta }^\alpha(\sigma)  +x^{\mu}(\sigma)  \bar{ \Psi %}_{\mu}^\alpha + 2\pi \bar{ \Psi }_{\mu}^\alpha N^\mu
%\Big) 
%( f^{ -1 } )_{ \alpha\alpha_1 }     C_\gamma^{\alpha_1 %\beta_1}      ( f^{ -1 } )_{ \beta_1\beta } \Psi_{\rho_1}^\beta %\big( \delta_{\nu_1}^\rho \delta_{\nu_2}^\gamma + %\delta_{\nu_2}^\rho \delta_{\nu_1}^\gamma \big)\,
+\frac{\pi }{k}N^\mu
 \bar{ \Psi }_{\mu_1}^\alpha
( f^{ -1 } )_{ \alpha\alpha_1 }     C_{\mu_2}^{\alpha_1 \beta_1}      ( f^{ -1 } )_{ \beta_1\beta } \Psi_{\mu_3}^\beta [
\delta^{\mu_1}_{\mu}\delta^{\mu_2}_{\nu_1}\delta^{\mu_3}_{\nu_2}
-\delta^{\mu_1}_{\nu_2}\delta^{\mu_2}_{\nu_1}\delta^{\mu_3}_{\mu} +
\delta^{\mu_1}_{\mu}\delta^{\mu_2}_{\nu_2}\delta^{\mu_3}_{\nu_1}
-
\delta^{\mu_1}_{\nu_1}\delta^{\mu_2}_{\nu_2}\delta^{\mu_3}_{\mu} ]\, .
\end{aligned}
\end{gathered}
\end{align}
Here we used fact that $H(2\pi) = 1$, while $N^\rho$ is winding number around compactified coordinate defined as

\begin{equation}
x^\mu(\sigma + 2 \pi) - x^\mu(\sigma) = 2\pi N^\mu.
\end{equation}

From this relation we can see that if we choose $x^\mu(\sigma) = 0$ than Poisson bracket has linear dependence on winding number. In cases where we do not have any winding number, we still have non-commutativity that is proportional to background fields.

Using the expression for sigma derivative of $y_\nu$ (\ref{sigma derivative of y}) and expression for Poisson bracket of T-dual coordinates (\ref{Poisson brackets T-dual}), we can find non-associative relations. Procedure is the same as for finding Poisson brackets of T-dual theory, we find Poisson bracket of sigma derivative and integrate with respect to sigma coordinate, this time integration is done once. Going along with this procedure we have the final result

\begin{align}
\begin{gathered}
\{  y_\nu (\sigma)  ,  \{ y_{\nu_1} (\sigma_1),  y_{\nu_2} (\sigma_2)  \} \} \cong \frac{1}{2k} H(\sigma_1-\sigma_2) \bar{ \Psi }_{\mu_1}^\alpha
( f^{ -1 } )_{ \alpha\alpha_1 }     C_{\mu_2}^{\alpha_1 \beta_1}      ( f^{ -1 } )_{ \beta_1\beta } \Psi_{\mu_3}^\beta \\
%%%%%%%%%%%%%%%%%%%%%%%%%%%%%%%%%%%%%%%%%%%%%%%%%%%%%%%%%%%%%%
\begin{aligned}
%&-\frac{2}{k^2} \big( \delta_{\nu_1}^\rho \delta_{\nu_2}^\gamma %+ \delta_{\nu_2}^\rho \delta_{\nu_1}^\gamma \big) \bar{ \Psi %}_{\rho}^\alpha
%( f^{ -1 } )_{ \alpha\alpha_1 }     C_\gamma^{\alpha_1 %\beta_1}      ( f^{ -1 } )_{ \beta_1\beta } \Psi_{\nu}^\beta %H(\sigma - \sigma_1) H(\sigma - \sigma_2)\\
%&+ \frac{2}{k^2}\bar{ \Psi }_{\nu}
%( f^{ -1 } )_{ \alpha\alpha_1 }     C_\gamma^{\alpha_1 %\beta_1}      ( f^{ -1 } )_{ \beta_1\beta } \Psi_{\rho_1}^\beta %\big( \delta_{\nu_1}^\rho \delta_{\nu_2}^\gamma + %\delta_{\nu_2}^\rho \delta_{\nu_1}^\gamma \big)H(\sigma - %\sigma_1) H(\sigma - \sigma_2).
%%%%%%%%%%%%%%%%%%%%%%%%%%%%%%%%%%%%%%%%%%%%%%%%%%%%%%%%%%%%%%%
\times \Big[\Big.
H(\sigma_1 - \sigma) [2\delta^{\mu_1}_{\nu}\delta^{\mu_2}_{\nu_2}\delta^{\mu_3}_{\nu_1} - 2 \delta^{\mu_1}_{\nu_1}\delta^{\mu_2}_{\nu_2}\delta^{\mu_3}_{\nu} - \delta^{\mu_1}_{\nu}\delta^{\mu_2}_{\nu_1}\delta^{\mu_3}_{\nu_2} + \delta^{\mu_1}_{\nu_2}\delta^{\mu_2}_{\nu_1}\delta^{\mu_3}_{\nu}     ]  \\
+H(\sigma_2 - \sigma)[2\delta^{\mu_1}_{\nu}\delta^{\mu_2}_{\nu_1}\delta^{\mu_3}_{\nu_2} - 2 \delta^{\mu_1}_{\nu_2}\delta^{\mu_2}_{\nu_1}\delta^{\mu_3}_{\nu} - \delta^{\mu_1}_{\nu}\delta^{\mu_2}_{\nu_2}\delta^{\mu_3}_{\nu_1} + \delta^{\mu_1}_{\nu_1}\delta^{\mu_2}_{\nu_2}\delta^{\mu_3}_{\nu}     ] \Big.\Big]\, .
\end{aligned}
\end{gathered}
\end{align}  

Since Jacobi identity is non-zero for T-dual theory we have that coordinate dependent RR field produces non-associative theory. However putting $\sigma=\sigma_2=\bar{\sigma}$ and $\sigma_1= \bar{\sigma} +2\pi$ we have following Jacobi identity 

\begin{align}
\begin{gathered}
\{  y_\nu (\bar{\sigma})  ,  \{ y_{\nu_1} (\bar{\sigma}+2 \pi),  y_{\nu_2} (\bar{\sigma})  \} \} \cong\\
\begin{aligned}
%%%%%%%%%%%%%%%%%%%%%%%%%%%%%%%%%%%%%%%%%%%%%%%%%%%%%%%%%%%%%5
%&-\frac{2}{k^2} \big( \delta_{\nu_1}^\rho \delta_{\nu_2}^\gamma %+ \delta_{\nu_2}^\rho \delta_{\nu_1}^\gamma \big) \bar{ \Psi %}_{\rho}^\alpha
%( f^{ -1 } )_{ \alpha\alpha_1 }     C_\gamma^{\alpha_1 %\beta_1}      ( f^{ -1 } )_{ \beta_1\beta } \Psi_{\nu}^\beta \\
%&+ \frac{2}{k^2}\bar{ \Psi }_{\nu}
%( f^{ -1 } )_{ \alpha\alpha_1 }     C_\gamma^{\alpha_1 %\beta_1}      ( f^{ -1 } )_{ \beta_1\beta } \Psi_{\rho_1}^\beta %\big( \delta_{\nu_1}^\rho \delta_{\nu_2}^\gamma + %\delta_{\nu_2}^\rho \delta_{\nu_1}^\gamma \big).
%%%%%%%%%%%%%%%%%%%%%%%%%%%%%%%%%%%%%%%%%%%%%%%%%%%%%%%%%%%%%%%%
\bar{ \Psi }_{\mu_1}^\alpha
( f^{ -1 } )_{ \alpha\alpha_1 }     C_{\mu_2}^{\alpha_1 \beta_1}      ( f^{ -1 } )_{ \beta_1\beta } \Psi_{\mu_3}^\beta [2\delta^{\mu_1}_{\nu}\delta^{\mu_2}_{\nu_2}\delta^{\mu_3}_{\nu_1} - 2 \delta^{\mu_1}_{\nu_1}\delta^{\mu_2}_{\nu_2}\delta^{\mu_3}_{\nu} - \delta^{\mu_1}_{\nu}\delta^{\mu_2}_{\nu_1}\delta^{\mu_3}_{\nu_2} + \delta^{\mu_1}_{\nu_2}\delta^{\mu_2}_{\nu_1}\delta^{\mu_3}_{\nu}     ]\, .
\end{aligned}
\end{gathered}
\end{align}

Examining equation (\ref{sigma derivative of y}), we notice that $\partial_{\sigma} y_\mu$ is not only a linear combination of initial coordinate and its momenta but also has terms that are proportional to fermionic coordinates. This might lead us to believe that T-dual theory would have nontrivial Poisson bracket between T-dual coordinate and fermionic coordinates. However, this is not the case, and it can be directly calculated by finding Poisson bracket between sigma derivative of T-dual coordinate and fermion coordinates (more details in \ref{appendix B}).  

\begin{equation}\label{eq:xteta}
\{ \theta^\alpha (\sigma),  y_{\mu} (\bar{\sigma})  \} \cong 0, \qquad \{ \bar{ \theta }^\alpha (\sigma),  y_{\mu} (\bar{\sigma})  \} \cong 0.
\end{equation}

%%%%%%%%%%%%%%%%%%%%%%%%%%%%%%%%%%%%%%%%%%%%%%%%%%%%%%%%%%%%%%%%%%%%%%%%%%%%%%%%%%%%%%%%%%%%%%%%%%%%%%%%%%%%%%%%%%%%%%%%%%%%%%%%%%%%%%%%%%%%%%%%%%%%%%%%%%%%%%%%%%%%%%%%%%%%%%%%%%%%%%%%%%%%%%%%%%%
\section{Conclusion}
\setcounter{equation}{0}

In this article we examined type II superstring propagating in presence of coordinate dependent RR field. This choice of background was in accordance with consistency conditions for background field and all calculations were made in approximation that are linear with respect to the space-time derivative of the  RR field, $C_\mu^{\alpha\beta}$, which is infinitesimal one. We have also excluded parts that were non-linear in fermionic coordinates and neglected pure spinor actions. Using equations of motion for fermionic momenta we obtained action that was expressed in terms of bosonic coordinates, their derivatives and derivatives of fermionic coordinates.

Action with our choice of background fields possessed translation symmetry, therefore we use generalized Buscher procedure that was developed for such cases. By substituting starting action with auxiliary action we gave up on locality in order to be able to find T-dual theory. Finding equations of motion of newly introduced Lagrange multipliers we were able to salvage starting action giving us assurance that auxiliary action we selected would produce correct T-dual theory. After this we found equations of motion for gauge fields and by inserting them into action, we found T-dual theory. 

Having found T-dual theory, we applied T-dual procedure once again as a more thorough way of checking if action we obtained was in fact correct T-dual of starting action. Unlike starting action, T-dual action was non-local from the start by virtue of containing $V^{(0)}$ term. Applying steps of generalized Buscher procedure we obtained starting action, again confirming that our choice of auxiliary action was correct.

We obtained non-commutativity relations in context of T-dual theory, where we used T-dual transformation laws as a bridge between Poisson brackets of starting theory and T-dual theory. T-dual transformation laws were expressed in terms of coordinates and momenta of original theory, which produced non-commutativity in T-dual theory. From expression for Poisson brackets (\ref{Poisson brackets T-dual}) we can see that non-commutativity is proportional to infinitesimal part of RR field. Non-commutativity relations are zero in case when $\sigma = \bar{\sigma}$, while in case where $\sigma = \bar{\sigma} + 2\pi$ we see the emergence of winding numbers. Noncommutativity parameters are linearly dependent on bosonic coordinates $x^\mu$ as well as on fermionic ones, $\theta^\alpha$ and $\bar\theta^\alpha$.

Taking into account Poisson brackets of T-dual coordinates and expression for sigma derivative of T-dual coordinate we were able to find non-associative relation for T-dual theory. In general case this relation was non-zero and it was proportional to infinitesimal constant, which is proportional to $C^{\alpha\beta}_\mu$. In special case when we put  $\sigma_1=\sigma_2=\bar{\sigma}$ and $\sigma_3= \bar{\sigma} +2\pi$ we noticed that non-associativity relation remains constant.

It should be noted that since we did not preform T-dualization along fermionic coordinates their Poisson structure would remain the same as in original theory. However, unlike original theory, T-dual coordinates depend on sigma derivatives of fermionic coordinates. This dependence does not affect the Poisson brackets of the T-dual coordinates and fermionic coordinates (\ref{eq:xteta}). So, T-dual SUSY algebra has non zero Poisson bracket of the bosonic coordinates, while the rest ones are zero. In further investigation we will study fermionic T-dualization and we expect the effect on the algebra of the fermionic coordinates. 
%%%%%%%%%%%%%%%%%%%%%%%%%%%%%%%%%%%%%%%%%%%%%%%%%%%%%%%%%%%%%%%%%%%%%%%%%%%%%%%%%%%%%%%%%%%%%%%%%%%%%%%%%%%%%%%%%%%%%%%%%%%%%%%%%%%%%%%%%%%%%%%%%%%%%%%%%%%%%%%%%%%%%%%%%%%%%%%%%%%%%%%%%%%%%%%%

\appendix
\setcounter{equation}{0}
\section{Obtaining $\beta^{\pm}_\mu$ terms}
\label{appendix A}

In this paper function $\beta^\pm_{\mu}(V)$ emerged in T-dual transformation laws as a consequence of variation of term that was proportional to $\Delta V$. Here we will present derivation of this function. 

Here we will use substitutions  $\partial_+ \bar{\Theta}^\alpha =\partial_+\bar{ \theta }^\alpha     +     v_+^{\nu_1}     \bar{ \Psi }_{\nu_1} ^\alpha $, $\partial_- \Theta^\beta = \partial_-\theta^\beta   + \Psi_{\nu_2} ^\beta  v_-^{\nu_2}$, also we will use $F_{\alpha \beta \rho}$ to represent term containing infinitesimal constant

\begin{align} \label{beta f}
\begin{gathered}
\int_{\Sigma} d^2\xi \partial_+ \bar{\Theta}^\alpha F_{\alpha \beta \rho} \Delta V^{(0) \rho} \partial_- \Theta^\beta = \int_{\Sigma} d^2\xi \epsilon^{m n} \partial_m \bar{\Theta}^\alpha F_{\alpha \beta \rho} \Delta V^{(0) \rho} \partial_n \Theta^\beta\\
\begin{aligned}
&=\int_{\Sigma} d^2\xi \Big[\frac{1}{2} \epsilon^{m n} \partial_m \bar{\Theta}^\alpha F_{\alpha \beta \rho} \Delta V^{(0) \rho} \partial_n \Theta^\beta - \frac{1}{2} \epsilon^{m n} \partial_n \bar{\Theta}^\alpha F_{\alpha \beta \rho} \Delta V^{(0) \rho} \partial_m \Theta^\beta\Big]\\
&= -\frac{1}{2}\int_{\Sigma} d^2\xi \Big[\epsilon^{m n}  \bar{\Theta}^\alpha F_{\alpha \beta \rho} \partial_m\Delta V^{(0) \rho} \partial_n \Theta^\beta
+\frac{1}{2} \epsilon^{m n} \partial_n \bar{\Theta}^\alpha F_{\alpha \beta \rho} \partial_m\Delta V^{(0) \rho}  \Theta^\beta\Big]\\
&=-\frac{1}{2} \int_{\Sigma} d^2\xi \epsilon^{m n}\partial_m\Delta V^{(0) \rho} \Big[
\bar{\Theta}^\alpha F_{\alpha \beta \rho} \partial_n \Theta^\beta
-\partial_n \bar{\Theta}^\alpha F_{\alpha \beta \rho}  \Theta^\beta
\Big]\\
&=-\frac{1}{2} \int_{\Sigma} d^2\xi\epsilon^{m n}v_m^\rho \Big[
\bar{\Theta}^\alpha F_{\alpha \beta \rho} \partial_n \Theta^\beta
-\partial_n \bar{\Theta}^\alpha F_{\alpha \beta \rho}  \Theta^\beta
\Big] = \int_{\Sigma} d^2 \xi v_m^\rho \beta^m_\rho.
\end{aligned}
\end{gathered}
\end{align}

Variation with respect to gauge field $v_\pm^\rho$, and setting $F_{\alpha \beta \rho} = -( f^{ -1 } )_{ \alpha\alpha_1 }     C_\mu^{\alpha_1 \beta_1}      ( f^{ -1 } )_{ \beta_1\beta }$  produces desired $\beta^\pm_\rho$ functions (\ref{beta 1}), (\ref{beta 2}) in equations of motion (\ref{equations of motion of v}). 
Here we have used the property that $( f^{ -1 } )_{ \alpha\alpha_1 }     C_\mu^{\alpha_1 \beta_1}      ( f^{ -1 } )_{ \beta_1\beta }$, ie. $F_{\alpha \beta \rho}$, is antisymmetric under exchange of $\alpha$ and $\beta$, this, in combination with the fact that we can express $\partial_+ \bar{\Theta} \partial_- \Theta$ as $\epsilon^{n m}\partial_n \bar{\Theta} \partial_m \Theta$, removes all terms proportional to $\partial_+ \partial_-$, using identity $\epsilon^{mn}\bar\Theta \partial_m\partial_n \Theta=0$.

It should be noted that $\beta_{\pm\mu}(V)$  functions are not unique, we could have obtained different function simply by not using using symmetrization in (\ref{beta f}). In case of non-symmetric $\beta_{\pm\mu}(V)$, all results that have been obtained would take a simpler form. We have chosen to work with symmetric function because results that are deduced from this case can be easily reduced, by neglecting terms, to simpler case.

\section{Poisson bracket between sigma derivatives of T-dual coordinate}
\label{appendix B}
\setcounter{equation}{0}

In this article, in order to find Poisson brackets of the T-dual coordinates we had to find  first Poisson brackets of the sigma derivative of T-dual coordinates, $y'_\mu\equiv \partial_\sigma y_\mu(\sigma)$. In this section we will demonstrate how to obtain Poisson brackets from Poisson brackets that contain sigma derivatives. We will use canonical form of the T-dual transformation law (\ref{sigma derivative of y}) and standard Poisson algebra, because the initial theory is geometric one. First, we have to calculate the following Poisson bracket

\begin{align}\label{pomocna jna1}
\begin{gathered}
\{ \partial_{\sigma_1}y_{\nu_1} (\sigma_1) , \partial_{\sigma_2} y_{\nu_2} (\sigma_2) \}= \frac{1}{2k}\Big[K_{\nu_2 \nu_1}(\sigma_2) \partial_{\sigma_2}\delta(\sigma_2 - \sigma_1) \Big.\\ 
\begin{aligned}
\Big. - K_{\nu_1 \nu_2} (\sigma_1) \partial_{\sigma_1} \delta(\sigma_1 - \sigma_2) + \partial_{\sigma_1} K_{\nu_1 \nu_2} (\sigma_1) \delta(\sigma_1 - \sigma_2) - \partial_{\sigma_2} K_{\nu_2 \nu_1} (\sigma_2) \delta(\sigma_2 - \sigma_1) \Big],
\end{aligned} 
\end{gathered}
\end{align}
where $K_{\mu \nu} (\sigma)$ is given by (\ref{Shortcut}).\\
On the other side we have

\begin{align}\label{pomocna jna2}
\begin{gathered}
\{\Delta y_{\nu_1}(\sigma_0,\sigma),\Delta y_{\nu_2}(\bar\sigma_0,\bar\sigma)\}=\int_{\sigma_0}^{\sigma} d \sigma_1 \int_{\bar{\sigma}_0}^{\bar{\sigma}} d \sigma_2 \{ \partial_{\sigma_1}y_{\nu_1} (\sigma_1) , \partial_{\sigma_2} y_{\nu_2} (\sigma_2) \}=\\
\{   y_{\nu_1} (\sigma), y_{\nu_2} (\bar{\sigma})   \} - \{   y_{\nu_1} (\sigma), y_{\nu_2} (\bar{\sigma}_0)   \} - \{   y_{\nu_1} (\sigma_0), y_{\nu_2} (\bar{\sigma})   \} + \{   y_{\nu_1} (\sigma_0), y_{\nu_2} (\bar{\sigma}_0)  \}\, ,\end{gathered}
\end{align}
where 
\begin{equation}
\Delta y_\mu(\sigma_0,\sigma)\equiv \int_{\sigma_0}^\sigma d\sigma_1 \partial_{\sigma_1} y_\mu(\sigma_1)=y_\mu(\sigma)-y_\mu(\sigma_0)\, .
\end{equation}

Combining the equations (\ref{pomocna jna1}) and (\ref{pomocna jna2}) we have 

\begin{align}
\begin{gathered}
\{\Delta y_{\nu_1}(\sigma_0,\sigma),\Delta y_{\nu_2}(\bar\sigma_0,\bar\sigma)\}=\frac{1}{2k} \int_{\sigma_0}^{\sigma} d \sigma_1 \int_{\bar{\sigma}_0}^{\bar{\sigma}} d \sigma_2 \Big[K_{\nu_2 \nu_1}(\sigma_2) \partial_{\sigma_2}\delta(\sigma_2 - \sigma_1) \Big. \\
\Big. - K_{\nu_1 \nu_2} (\sigma_1) \partial_{\sigma_1} \delta(\sigma_1 - \sigma_2) + \partial_{\sigma_1} K_{\nu_1 \nu_2} (\sigma_1) \delta(\sigma_1 - \sigma_2) - \partial_{\sigma_2} K_{\nu_2 \nu_1} (\sigma_2) \delta(\sigma_2 - \sigma_1) \Big].
\end{gathered}
\end{align}
By applying partial integration, it is straightforward to extract the Poisson bracket of T-dual coordinates given by (\ref{Poisson brackets T-dual}). 

In paper \cite{Generalized Buscher procedure 3} it has been shown that Poisson brackets between $\sigma$ derivatives of coordinates have following form

\begin{equation} \label{c.1}
\{\partial_{\sigma_1} X_\mu (\sigma_1) , \partial_{\sigma_2} Y_\nu (\sigma_2 )\}\cong \partial_{\sigma_1} K_{\mu\nu}(\sigma_1) \delta(\sigma_1 - \sigma_2) + L_{\mu \nu} (\sigma_1) \partial_{\sigma_1} \delta (\sigma_1 - \sigma_2).
\end{equation}
Applying integrating twice and using partial integration this equation reduces to

\begin{equation}
\{ X_\mu (\sigma_1), Y_\nu (\sigma_2)  \} \cong - \big[ K_{\mu \nu} (\sigma_1) - K_{\mu \nu} (\sigma_2) + L_{\mu \nu} (\sigma_2) \big] \delta(\sigma_1 - \sigma_2).
\end{equation}
In our case, we can bring equation (\ref{pomocna jna1}) to the form of equation (\ref{c.1}) by making following substitutions

\begin{equation}
\begin{gathered}
\partial_{\sigma_1}K_{\nu_1 \nu_2} (\sigma_1) = \partial_{\sigma_1}K_{\nu_1 \nu_2} (\sigma_1), \qquad \partial_{\sigma_2}K_{\nu_2 \nu_1} (\sigma_2) = \partial_{\sigma_2}K_{\nu_2 \nu_1} (\sigma_2),\\
K_{\nu_1 \nu_2}({\sigma_1}) = -L_{\nu_1 \nu_2} ({\sigma_1}), \qquad K_{\nu_2 \nu_1}({\sigma_2}) = -L_{\nu_2 \nu_1} ({\sigma_2}).
\end{gathered}
\end{equation}
Because we chose to work with symmetric $\beta^\pm_\mu$ function we obtain duplicated terms in (\ref{c.1}).

Same procedure can be applied to find Poisson bracket between T-dual coordinate and fermionic momenta. That is, we start from Poisson bracket for sigma derivative of T-dual coordinate and fermionic momenta, then integrate once and compare left and right hand sides.

The step function $H(x)$ is defined as
\begin{equation}\label{eq:fdelt}
H(x)=\int_0^x ds \delta(s)=\frac{1}{2\pi}\Sigma_{n\in Z}\int_0^x e^{ins}=\left\{\begin{array}{ll}
0 & \textrm{if $x=0$}\\
1/2 & \textrm{if $0<x<2\pi$}\, .\\
1 & \textrm{if $x=2\pi$} \end{array}\right .
\end{equation}

\end{document}